\documentclass[twocolumn,pra,aps]{revtex4}
\usepackage{graphicx}

\def\duzomniejsze{<\kern-.7mm<}
\def\duzowieksze{>\kern-.7mm>}

\def\textbf#1{{\bf #1}}
\def\beq{\begin{equation}}
\def\eeq{\end{equation}}
\def\be{\begin{equation}}
\def\ee{\end{equation}}
\def\ben{\begin{eqnarray}}
\def\een{\end{eqnarray}}
\def\beqa{\begin{eqnarray}}
\def\eeqa{\end{eqnarray}}
\def\eea{\end{array}}
\def\bea{\begin{array}}
\newcommand{\bei}{\begin{itemize}}
\newcommand{\eei}{\end{itemize}}
\newcommand{\bee}{\begin{enumerate}}
\newcommand{\eee}{\end{enumerate}}

\def\>{\rangle}
\def\<{\langle}

\begin{document}

\title{Competition of different evaluation schemes in the continuous variable game}

\begin{abstract}
An asymmetric generalization of classical Cournot's duopoly game was
introduced and the simulation scheme of its quantized version was
analyzed. In this scheme, the player assigned by a 'classical'
measurement scheme always wins the player assigned by a quantum
measurement scheme. It was shown that the fluctuation causes the
disadvantage game rule of the player measured by the quantum
apparatus in this specific case.

\end{abstract}
\author{Shang-Bin Li}\email{stephenli74@yahoo.com.cn}

\affiliation{Research and development department of AmBrite
optoelectronic (Kunshan) Ltd., Jingde road 28, kunshan, Suzhou, P.R.
China}

\maketitle

\section * {I. INTRODUCTION OF ASYMMETRIC GENERALIZATION OF CLASSICAL COURNOT's DUOPOLY}

Significant interest has been focused on quantum games
\cite{meyer,eisert,Enk,Enk2002,Benjamin2001prl,Flitney2007pla,Du2002,Benjamin2001,Iqbal,Iqbal2004,cabello,f},
a new born branch of quantum information theory, which can exploit
both quantum superposition \cite{meyer,Enk} and quantum entanglement
\cite{eisert,Enk2002}. Among them, asymmetric quantum games have
also been investigated \cite{Du2003,Li2006,Flitney2003}. Some
previous studies on the asymmetric quantum games have revealed the
player utilizing quantum strategies has the advantage than the one
only using classical strategies \cite{Flitney2003}. The role of
quantum correlation or classical correlation on the quantum
Prisoner's dilemma has also been investigated \cite{Shimamura}, and
the influence of quantum fluctuations on quantum games has been
discussed \cite{Guinea}. However, there is still very little
attention focusing on the asymmetric quantum games in which the
asymmetry is caused by different measurement schemes or evaluation
schemes assigned to different players.

Recently, a simulation scheme without any entanglement involved of
quantized Cournot's Duopoly has been presented\cite{Li2011}, which
is different from the quantization scheme containing the
intermediate entanglement in Ref.\cite{Li2002}. It has been shown
that the scheme using classical measuring apparatus is advantage to
the one using the quantum measuring apparatus \cite{Li2011}. Here,
we will analyze its asymmetric version in which two firms are
assigned as different measurement schemes or evaluation schemes.
Firstly, let us introduce an asymmetric generalization of classical
Cournot's duopoly game \cite{c}, two firms simultaneously decide the
quantities $q_1$ ($q_1\geq0$) and $q_2$ ($q_2\geq0$) respectively,
of a homogeneous product released on the market. Two firms have
different execution precisions for their strategies. Assuming the
firm 1 can definitely execute its strategy without any possible
deviation, and the firm 2 can only execute its strategy with
outcomes $\eta_2\in{N}$ probabilistically distributing on the set of
the nonnegative integer with the probability distribution function
$D(q_2,\eta_2)$ which has the constraint condition
$\sum^{\infty}_{\eta_2=0}D(q_2,\eta_2)=1$ and the expected value
$q_2$ given by $q_2=\sum^{\infty}_{\eta_2=0}\eta_2D(q_2,\eta_2)$.
The distribution $D(q_2,\eta_2)$ of the firm 2 is common knowledge.
Suppose $\Lambda$ is the total quantity, i.e., $\Lambda=q_1+\eta_2$,
and the market-clearing price is given by $P(\Lambda)=a-\Lambda$ for
$\Lambda\leq{a}$ and $P(\Lambda)=0$ for $\Lambda>{a}$. The unit cost
of producing the product is assumed to be $c$ with $c<a$. In the
extreme case with $a\rightarrow\infty$ and $c\rightarrow\infty$ but
keeping $k=a-c$ a finite nonnegative constant, the average payoff
functions of two firms can be obtained as \beqa
u_1(q_1,q_2)&=&\sum^{\infty}_{\eta_2=0}q_1[P(\Lambda)-c]D(q_2,\eta_2)\nonumber\\
&=&q_1[k-(q_1+q_2)]\nonumber\\
u_2(q_1,q_2)&=&\sum^{\infty}_{\eta_2=0}\eta_2[P(\Lambda)-c]D(q_2,\eta_2)\nonumber\\
&=&q_2[k-(q_1+q_2)]-\Delta^2\eta_2\eeqa where \be
\Delta\eta_2=\{[\sum^{\infty}_{\eta_2=0}\eta^2_2D(q_2,\eta_2)]-q^2_2\}^{\frac{1}{2}}
\ee is the standard deviation of $\eta_2$. For clarifying the role
of statistical phenomenon of the optical field in the later
simulation scheme of this kind of the game, adopting the Mandel-Q
parameter $Q(q_2)\equiv\frac{\Delta^2\eta_2}{q_2}-1$
\cite{kimble1977}, which is related with the second order intensity
correlation function $g^{(2)}(0)$ via
$g^{(2)}(0)=\frac{Q(q_2)+q_2}{q_2}$, the above average payoff
function can be rewritten as \beqa u_1(q_1,q_2)
&=&q_1[k-(q_1+q_2)]\nonumber\\
u_2(q_1,q_2)&=&q_2[k-(q_1+q_2+Q(q_2)+1)].\eeqa It explicitly shows
the asymmetry of this game increases with the Mandel-Q parameter.

In the above game, a strategy profile $\{q^{\ast}_1,q^{\ast}_2\}$ is
a Nash equilibrium if no unilateral deviation in strategy by firm 1
or 2 is profitable for firm 1 or 2, respectively, that are
$u_1(q^{\ast}_1,q^{\ast}_2)>u_1(q_1,q^{\ast}_2)$ and
$u_2(q^{\ast}_1,q^{\ast}_2)>u_1(q^{\ast}_1,q_2)$ holding for any
$q_1\neq{q}^{\ast}_1$ and $q_2\neq{q}^{\ast}_2$ \cite{osborne}.
Since $\frac{\partial^2{u}_1}{\partial{q}^2_1}=-2<0$ and
$\frac{\partial^2{u}_2}{\partial{q}^2_2}=-2-\frac{\partial^2{\Delta^2\eta_2}}{\partial{q}^2_2}$,
Nash equilibrium condition is
$\frac{\partial{u}_1}{\partial{q}_1}=\frac{\partial{u}_2}{\partial{q}_2}=0$
if the inequality
$\frac{\partial^2\Delta^2\eta_2}{{\partial{q}_2}^2}>-2$ holds for
the solution of equilibrium condition. Solving for the Nash
equilibrium yields the equilibria, \be
q^{\ast}_1=\frac{k-q^{\ast}_2}{2}, \ee where $q^{\ast}_2$ is the
nonnegative root of the equation \be
q^{\ast}_2=\max[\frac{k}{3}-\frac{2}{3}\frac{\partial\Delta^2\eta_2}{\partial{q}^{\ast}_2},0].\ee
Thereafter, it is assumed Eq.(5) has unique nonnegative root. In the
above derivations, it has been assumed $\Delta^2\eta_2$ is
differentiable and simultaneously
$\frac{\partial^2\Delta^2\eta_2}{{\partial{q^{\ast}}_2}^2}>-2$. One
will find the sign of
$\delta({q}^{\ast}_2)\equiv\frac{\partial\Delta^2\eta_2}{\partial{q}^{\ast}_2}$
plays a crucial role in the Nash equilibria, which determines the
degree of the strategic asymmetry in the game.

In what follows, let us consider some specific cases corresponding
to zero, negative, and positive values of $\delta({q}^{\ast}_2)$.
For example, in the case 1 with $\Delta^2(\eta_2)=const.$, we have
$\delta({q}^{\ast}_2)=0$, $q^{\ast}_1=q^{\ast}_2=\frac{k}{3}$, and
$u_1(q^{\ast}_1,q^{\ast}_2)=u_2(q^{\ast}_1,q^{\ast}_2)+\Delta^2(\eta_2)=\frac{k^2}{9}$.

For the case 2 with $\delta({q}^{\ast}_2)<0$ (namely the firm 2 can
improve its execution efficiency and precision with the increase of
the product manufactured), the firm 2 can obtain more profit than
the firm 1 if only \be
u_1(q^{\ast}_1,q^{\ast}_2)-u_2(q^{\ast}_1,q^{\ast}_2)=\frac{1}{3}\delta({q}^{\ast}_2)(k+\delta({q}^{\ast}_2))+\Delta^2(\eta_2)<0.
\ee

In the case 3, we assume the Poisson distribution
$D(q_2,\eta_2)=e^{-q_2}\frac{{q_2}^{\eta_2}}{\eta_2!}$. In this
case, $Q(q_2)=0$, $ q^{\ast}_1=\min[\frac{k+1}{3},\frac{k}{2}]$,
$q^{\ast}_2=\max[\frac{k-2}{3},0]$. At this equilibrium the payoffs
for the firm $i$ ($i=1,2$) are \beqa
u^{\ast}_1&=&\min[\frac{(k+1)^2}{9},\frac{k^2}{4}],\nonumber\\
u^{\ast}_2&=&\max{}^2[\frac{k-2}{3},0].\eeqa In the following
section, it will be shown that this equilibrium fails to be the
optimal solution for most values of the parameter $k$, because two
firms could simultaneously achieve more payoff than the one in
Eq.(7) from the region D of Fig.3.
\begin{figure}
\centerline{\includegraphics[width=2.0in]{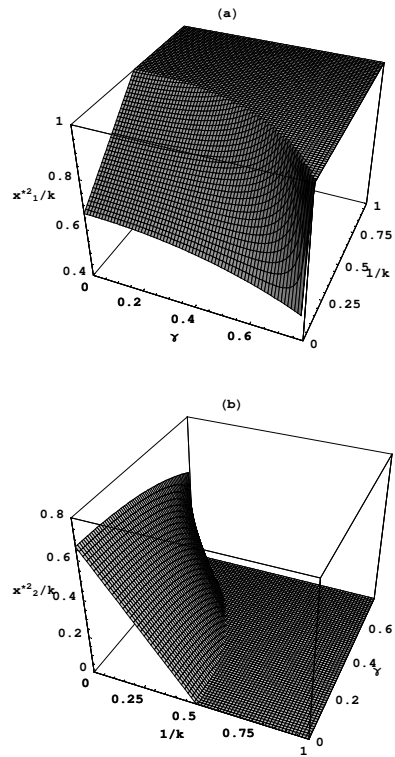}}
\caption{(a) The scaled Nash equilibrium strategy $x^{\ast2}_1/k$ of
the firm 1 and (b) the scaled Nash equilibrium strategy
$x^{\ast2}_2/k$ of the firm 2 at Nash equilibrium are plotted as the
function of $1/k$ and $\gamma$. When $\cos(2\gamma)=\frac{1}{k-1}$,
the transition-like behaviors occur.}
\end{figure}

In the above discussion, the different execution precision of the
strategies cause the asymmetry of the game. Equivalently, different
evaluation schemes for two firms may result in the same asymmetric
game. The above asymmetric generalization of classical Cournot's
Duopoly can not only be applied to economical and managemental
fields, may also find its application in a class of two "detector"
systems, such as the 3D vision system. For example, the
self-regulating aperture or pupils play the similar role as two
firms in this game which control the light intensity analoging with
the $q_i$ ($i=1,2$) in Eq.(1); CCD, CMOS or retina play the
evaluation role; The clarity or the signal noise ratio of the
picture maybe have monotonic relations with the payoffs in Eq.(1).
As an interesting illustration, there is no one owning the same left
and right eyes, and the brain of human being will control two
pupils, and simultaneously coordinate the information received by
left and right retina. Thus it is worthy to investigate how the
potential cooperation affects this kind of asymmetric game.

\section * {II. SIMULATION SCHEME OF QUANTIZED COURNOT's DUOPOLY WITH DIFFERENT MEASUREMENT APPARATUSES}

Considering a simulation scheme for the quantized version of above
asymmetric game of the case 3, two single-mode optical fields which
are initially at the vacuum state $|0\rangle_1\otimes|0\rangle_2$
are sent to firm 1 and firm 2, respectively. the strategic moves of
firm 1 and firm 2 are represented by the displacement operators
$\hat{D}_1$ and $\hat{D}_2$ locally acted on their individual
optical fields. The firms are restricted to choose their strategies
from the sets \be
S_i=\{\hat{D}_i(x_i)=\exp[\frac{\sqrt{2}}{2}x_i(a^{\dagger}_i-a_i)]|x_i\in[0,\infty)\},~i=1,2
\ee where $a_i$ and $a^{\dagger}_i$ are the annihilation and
creation operators of the $i$th mode optical field. In this stage,
the state of the game becomes a direct product of two coherent
states $|\frac{\sqrt{2}}{2}x_1\rangle$ and
$|\frac{\sqrt{2}}{2}x_2\rangle$, \be
|\psi\rangle=|\frac{\sqrt{2}}{2}x_1\rangle\otimes|\frac{\sqrt{2}}{2}x_2\rangle.
\ee Having executed their moves, firm 1 and firm 2 forward their
optical fields containing enough coherent pulses to the final
measurement, prior to which a beam splitter operation
$\hat{J}(\gamma)=\exp[i\gamma({a}^{\dagger}_1a_2+{a}^{\dagger}_2a_1)]$
($\gamma\in[0,\frac{\pi}{4})$) is carried out. Therefore the final
state prior to the measurement can be expressed as \beqa
|\Psi\rangle&=&|\frac{\sqrt{2}}{2}{x}_1\cos\gamma+\frac{\sqrt{2}}{2}i{x}_2\sin\gamma\rangle\nonumber\\
&&\otimes|\frac{\sqrt{2}}{2}{x}_2\cos\gamma+\frac{\sqrt{2}}{2}i{x}_1\sin\gamma\rangle.
\eeqa
\begin{figure}
\centerline{\includegraphics[width=2.0in]{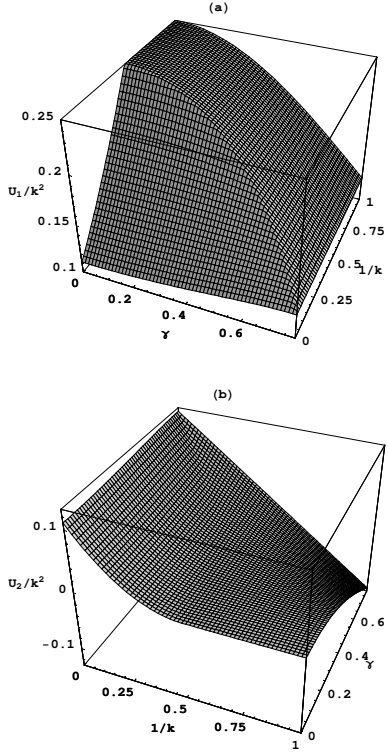}}
\caption{(a) The scaled payoff $U_1/k^2$ of the firm 1 and (b) the
scaled payoff $U_2/k^2$ of the firm 2 at Nash equilibrium are
plotted as the function of $1/k$ and $\gamma$.}
\end{figure}
Next the measurement on the photon number of the optical fields is
carried out, which is usually done by photon-detector. We focus on
the asymmetric game in which the judge uses different detecting
apparatuses to measure the final quantities of two firms.
"classical" measuring apparatus is assigned to firm 1, which can
only give out the expected values of the photon numbers of the
optical fields such as the optical power meter. Quantum measuring
apparatus is assigned to firm 2, such as the highly sensitive
quantum photon-counter which can measure the photon number and its
distribution of the quantum optical fields. For the final state in
Eq.(10), the expected value of the photon number of the firm 1 is
$n_1=\frac{1}{2}(x^2_1\cos^2\gamma+x^2_2\sin^2\gamma)$. The photon
number of the optical field of the firm 2 is the non-negative
integer $m_2$ with the Poisson probability distribution $P_{m_2}$
given by \beqa
P_{m_2}&=&e^{-\frac{1}{2}(x^2_2\cos^2\gamma+x^2_1\sin^2\gamma)}\nonumber\\
&&\frac{(\frac{1}{2}(x^2_2\cos^2\gamma+x^2_1\sin^2\gamma))^{m_2}}{m_2!}.
\eeqa. The average payoffs are given by \be
u^{Q}_i(\hat{D}_1,\hat{D}_2)=\langle{u_i(n_1,m_2)}\rangle, \ee where
\be
\langle{u_i(n_1,m_2)}\rangle=\sum^{\infty}_{m_2=0}u_i(n_1,m_2)P_{m_2}
\ee denotes the average of $u_i(n_1,m_2)$ taken over all possible
values of $m_2$ with the Poisson distribution $P_{m_2}$.
\begin{figure}
\centerline{\includegraphics[width=2.0in]{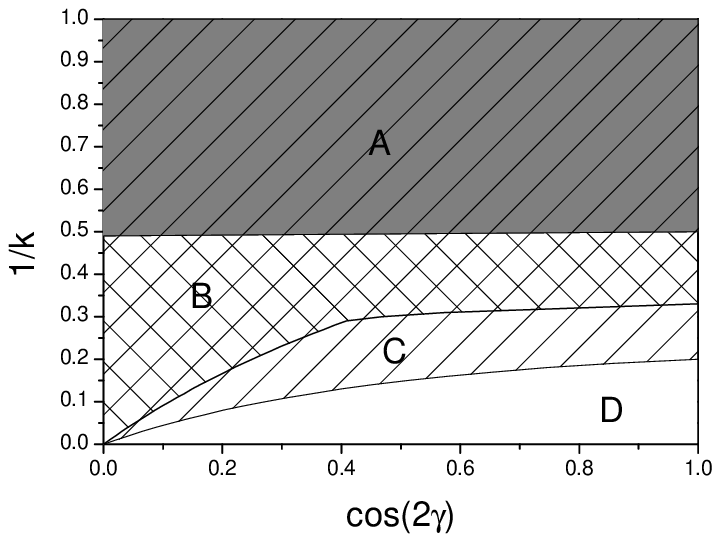}}
\caption{The distinct regions corresponding to the increase or
decrease of the individual payoffs of two firms at Nash equilibrium
and their total payoff. The payoff $U^{\ast}_1$ increases with
$\gamma$ in region D and decreases with $\gamma$ in combined region
A+B+C; The payoff $U^{\ast}_2$ increases with $\gamma$ in combined
regions B+C+D and decreases with $\gamma$ in region A; The total
payoff $U^{\ast}_1+U^{\ast}_2$ increases with $\gamma$ in region C+D
and decreases with $\gamma$ in combined region A+B.}
\end{figure}
It is still assumed $a$ and $c$ tend to infinity but keeping
$k=a-c\geq1$ a finite constant. Thus the quantum payoffs for two
firms are obtained as \beqa
u^{Q}_1(\hat{D}_1,\hat{D}_2)&=&\frac{1}{2}(x^2_1\cos^2\gamma+x^2_2\sin^2\gamma)[k-\frac{1}{2}(x^2_1+x^2_2)],\nonumber\\
u^{Q}_2(\hat{D}_1,\hat{D}_2)&=&\frac{1}{2}(x^2_2\cos^2\gamma+x^2_1\sin^2\gamma)[k-1-\frac{1}{2}(x^2_1+x^2_2)].
\eeqa Here, when $\hat{J}(\gamma)=I$ (the identity operator), the
scheme can return to the case 3 in the classical asymmetric
Cournot's Duopoly, in which quantum fluctuation of the optical field
causes the reduce of the payoffs of firm 2. Solving for the Nash
equilibrium yields the unique one \beqa
x^{\ast2}_1&=&\frac{2\cos^2(\gamma)[k+\sec(2\gamma)]}{2+\cos(2\gamma)}\nonumber\\
x^{\ast2}_2&=&\frac{2\cos^2(\gamma)[k-1-\sec(2\gamma)]}{2+\cos(2\gamma)}
\eeqa under the condition of $\cos(2\gamma)\geq\frac{1}{k-1}$, and
\beqa
x^{\ast2}_1&=&k,\nonumber\\
x^{\ast2}_2&=&0\eeqa under the condition of
$\cos(2\gamma)<\frac{1}{k-1}$. In Fig.1, the scaled Nash equilibrium
strategies of two firms are plotted as the function of $\gamma$ and
$1/k$. The profits of two firms in Nash equilibrium can be easily
derived by substituting Eq.(15) or Eq.(16) into Eq.(14). \beqa
U^{\ast}_1&=&\frac{\cos^2(\gamma)[1+2k+\cos(2\gamma)]^2}{4(2+\cos(2\gamma))^2}\nonumber\\
U^{\ast}_2&=&\frac{\cos^2(\gamma)[3-2k+\cos(2\gamma)]^2}{4(2+\cos(2\gamma))^2}
\eeqa when $\cos(2\gamma)\geq\frac{1}{k-1}$, and \beqa
U^{\ast}_1&=&\frac{1}{4}k^2\cos^2(\gamma),\nonumber\\
U^{\ast}_2&=&\frac{1}{4}k(k-2)\sin^2(\gamma)\eeqa when
$\cos(2\gamma)<\frac{1}{k-1}$. $U^{\ast}_1$ increases with $\gamma$
if both $k>5$ and $\cos(2\gamma)>\frac{2k-5-\sqrt{4k^2-20k+9}}{2}$
are simultaneously satisfied. Otherwise, $U^{\ast}_1$ decreases with
$\gamma$. $U^{\ast}_2$ increases with $\gamma$ if $k>2$ and
decreases with $\gamma$ if $1\leq{k}<2$. When $k=2$, the payoff of
the firm 2 at the Nash equilibrium is always zero. In Fig.2, the
scaled payoffs of two firms at Nash equilibrium are depicted. The
extraordinary aspects of the present asymmetric case are very
distinct from the original symmetric game. The payoffs are not only
related to the value of $\gamma$, but also dependent on the value of
$k$. A second-order transition-like behavior of the payoffs occurs
at the points satisfying $\cos(2\gamma)=\frac{1}{k-1}$, which
results from the asymmetry of the game. For the firm 1 assigned a
classical evaluation, its payoff at the Nash equilibrium is always
larger than the one of firm 2. As $k$ increases, the degree of
asymmetry decreases which causes the scaled payoffs $U_1/k^2$ and
$U_2/k^2$ to become closer to each other.
\begin{figure}
\centerline{\includegraphics[width=2.0in]{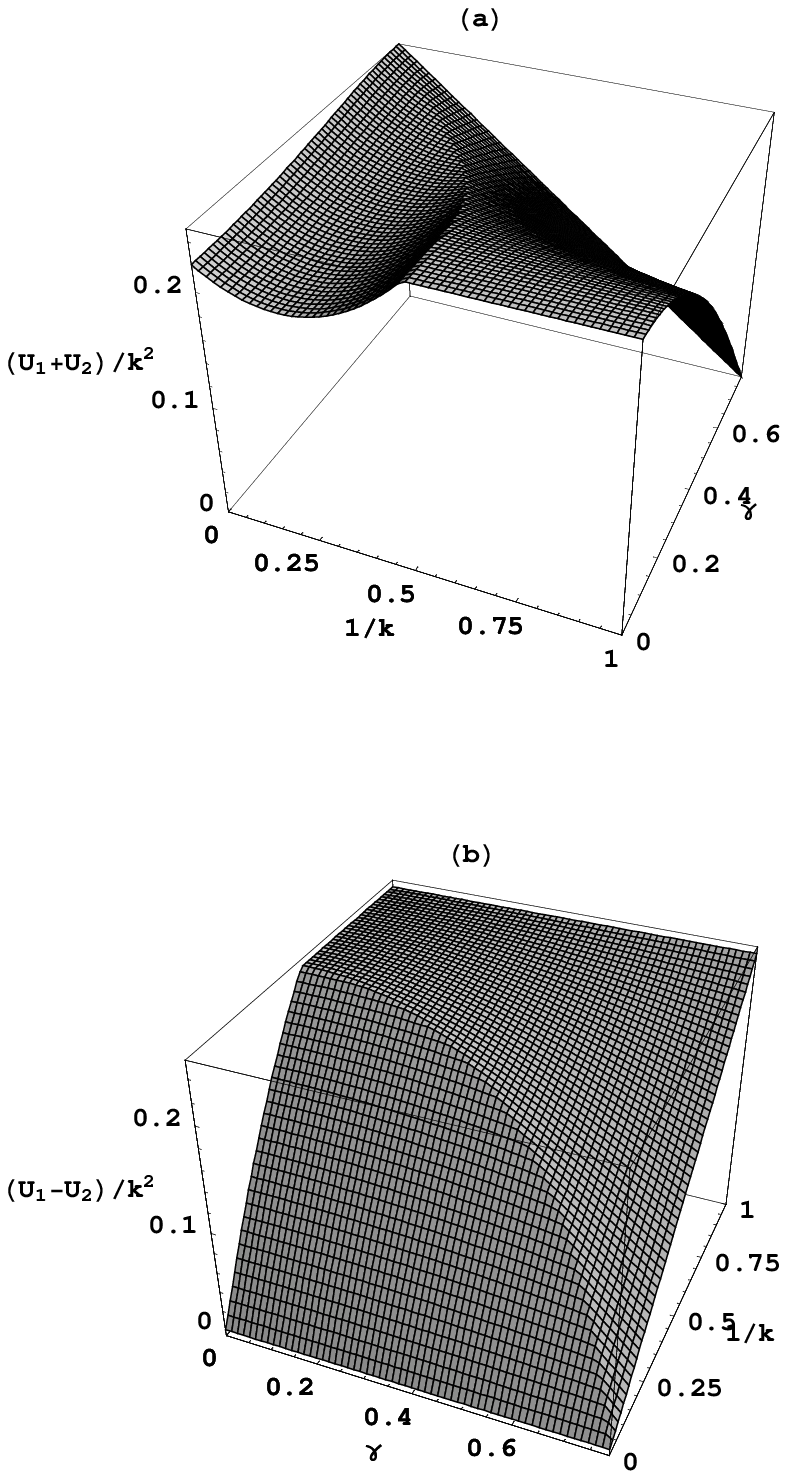}}
\caption{(a) The scaled total payoff $(U_1+U_2)/k^2$ of the firms
and (b) the scaled payoff difference $(U_1-U_2)/k^2$ of the firms at
Nash equilibrium are plotted as the function of $\gamma$ and $1/k$.}
\end{figure}
When $\cos(2\gamma)\geq\frac{1}{k-1}$, we have \beqa
U^{\ast}_1+U^{\ast}_2&=&\frac{\cos^2(\gamma)[11+8k(k-1)+8\cos(2\gamma)+\cos(4\gamma)]}{4(2+\cos(2\gamma))^2}\nonumber\\
U^{\ast}_1-U^{\ast}_2&=&\frac{\cos^2(\gamma)(2k-1)}{2+\cos(2\gamma)},
\eeqa and when $\cos(2\gamma)<\frac{1}{k-1}$, \beqa
U^{\ast}_1+U^{\ast}_2&=&\frac{1}{4}k(k-1+\cos(2\gamma))\nonumber\\
U^{\ast}_1-U^{\ast}_2&=&\frac{1}{4}k(1+(k-1)\cos(2\gamma)).\eeqa The
region in the plane of $\cos(2\gamma)$ and $1/k$ for the increase of
the total payoff has been clearly indicated in Fig.3. It is
interesting to quantificationally analyze how the asymmetry of the
evaluation schemes affects the payoff. For it, we need a rational
quantitative definition of the asymmetry, symmetry and cooperation
of the game. Assuming $s\equiv\min[\frac{2}{k},1]$ to be the degree
of asymmetry which varies from $1$ to 0 for $k\in[1,\infty)$, and
$\bar{s}=1-s$ to be the degree of symmetry.
$\xi\equiv\frac{1-\cos(2\gamma)}{1+\cos(2\gamma)}$ could be regarded
as the degree of cooperation. Then, the expression of the scaled
payoff difference can be rewritten as \be
(U^{\ast}_1-U^{\ast}_2)/k^2=\frac{1-\frac{1}{4}(1+\bar{s})^2}{3+\xi}
\ee when $\bar{s}>\xi$; Otherwise, when $0<\bar{s}<\xi$, the scaled
payoff difference is given by \be
(U^{\ast}_1-U^{\ast}_2)/k^2=\frac{1}{4}\frac{1-\bar{s}\xi}{1+\xi}.
\ee For $0=\bar{s}<\xi$, \be
(U^{\ast}_1-U^{\ast}_2)/k^2=\frac{1}{4}\frac{1+(2/k-1)\xi}{1+\xi}.
\ee At the second-order transition boundary labeled by
$\bar{s}=\xi$, \beqa
(U^{\ast}_1-U^{\ast}_2)/k^2&=&\frac{1}{4}s,\nonumber\\
(U^{\ast}_1+U^{\ast}_2)/k^2&=&\frac{1}{4}\frac{1+\bar{s}^2}{1+\bar{s}}.
\eeqa The sum and difference of the scaled payoffs of two firms are
plotted as the function of $\gamma$ and $1/k$ in Fig.4.

Finally, we briefly discuss the physical realization of present
scheme. In realistic situation, two players may initially have two
single-mode laser diodes which can radiate continuous coherent
laser. Two players modulate their laser intensity according to their
individual strategic moves and generate enough coherent pulses
sequence. Then the coherent pulse sequences pass a beam splitter and
are detected by the optical power meter and the high-efficient
quantum photon-detector, respectively. The optical power meter gives
out the average power of the coherent optical pulses. It is easy to
transform the average power to the average photon number contained
in each pulse. The high-efficient quantum photon-detector should
distinguish the Fock states and record their corresponding
probability distributions. Up to date, it is still very difficult to
experimentally implement multi-photon detection with high enough
efficiency. It has been demonstrated a system capable of directly
measuring the photon-number state of a single pulse of light using a
superconducting transition-edge sensor microcalorimeter. The
photon-number distribution of a weak pulsed-laser source at 1550 nm
has been verified \cite{Miller}. A charge integration photon
detector that enables the efficient measurement of photon number
states at 1530nm wavelengths with a quantum efficiency of $80$
percent has been also presented \cite{Fujiwara2005}.

\section * {III. CONCLUDING REMARKS}

In summary, we presented an asymmetric generalization of classical
Cournot's duopoly game, and proposed the simulation scheme of its
quantized version, in which the player assigned by a 'classical'
measurement scheme always wins the player assigned by a quantum
measurement scheme. Due to the common affection of the asymmetric
evaluation scheme, the cooperation induced by the beam splitter, and
the marginal effect of strategic space, there exists a second-order
transition-like behaviors of the payoffs at Nash equilibrium. At the
transition boundary, the symmetry and the cooperation become
balanced, and the scaled payoff difference is proportional to the
degree of asymmetry of the game.

This simulation scheme is symmetric photon-loss free, namely the
symmetric photon-loss does not alter the unique property of this
quantized asymmetric Cournot's duopoly. If both two firms have the
complete information about the photon loss, they can adjust their
strategies according to the transformation
$x_i\rightarrow{x}_ie^{\frac{\kappa}{2}t}$ ($\kappa$ denotes the
photon loss rate), which can guarantee the final payoffs are
invariant under the influence of the photon loss process.

The Mandel Q parameter of the optical field heavily affects the
characteristics of this kind of the games. In the noncooperation
case with $\xi=0$, if two players adopt certain kind of optical
field fulfilling the inequality (6) as the carrier of their
strategic moves, the player assigned by high-efficient quantum
photon-detector have the chance to win. While for $\xi\neq0$, the
cases become more complicated and will be analyzed in future work.

\section * { ACKNOWLEDGMENTS}

SL thanks the anonymous ones for helpful comments.

\bibliographystyle{apsrev}

\end{document}